     \documentstyle[12pt]{article}   
     \newlength{\dinwidth}                       
     \newlength{\dinmargin}                      
\def\Journal#1#2#3#4{{#1} {\bf #2}, #3 (#4)}

\def\NPB{{\em Nucl. Phys.} B}

\def\PRD{{\em Phys. Rev.} D}
\def\ZPC{{\em Z. Phys.} C}
\def\lsim{\mathrel{\rlap{\lower4pt\hbox{\hskip1pt$\sim$}}
    \raise1pt\hbox{$<$}}}                % less than or approx. symbol
\def\gsim{\mathrel{\rlap{\lower4pt\hbox{\hskip1pt$\sim$}}
    \raise1pt\hbox{$>$}}}                % greater than or approx. symbol
\begin{document}
\input{epsf.tex}
\vspace*{10mm}
\begin{center}  \begin{Large} \begin{bf}
 QED Radiative Effects in Diffractive \\ 
 Vector Meson  Electro- and Photoproduction \\
 at HERA Energies\\
  \end{bf}  \end{Large}
  \vspace*{5mm}
  \begin{large}
  I.Akushevich\\
  \end{large}
\end{center}
National Center of  Particle and  High Energy Physics, Minsk Belarus, e-mail:
aku@hep.by
\begin{quotation}
\noindent
{\bf Abstract:}
QED radiative corrections to the cross section of diffractive electro- and
photoproduction of vector mesons is calculated at
HERA energies. 
Both semi-analytical
and Monte Carlo approaches are discussed and
compared.
\end{quotation}

\section{Introduction}\label{intro}

The experience with data analysis in inclusive measurements of deep
inelastic scattering (DIS) shows
that radiative effects are very important and should be taken into
account with maximally possible accuracy. Numerical results
obtained for the case of the
HERA collider \cite{BSpi1,BSpi2,Bar1,Bar2,Spi,Bar,ASpi,AISh}
have shown that the ratio of the observed and the Born cross sections
varies 
from 0.5 to 2.0, and 
even larger values are reached at the borders of the kinematical
region. The calculation of radiative corrections (RC) 
for semi-inclusive and exclusive
reactions can not be easily reduced to the case of inclusive
scattering, but rather requires separate
considerations. There are two important approaches which can be applied
for this task. The first one 
is the 
so-called semi-analytical approach, where covariant formulae for RC are 
obtained after applying the procedure of covariant cancellation of
infrared divergences \cite{BSh} and integration over the photon phase
space. 
No other approximation than the ultrarelativistic one is used within
this 
approach. Explicit analytical formulae for RC in 
diffractive
vector meson electroproduction were obtained in Ref.\ \cite{Aku}. The
radiative correction factor has the following form   
\begin{equation}
\eta=\frac{\sigma_{obs}}{\sigma_0}=\exp(\delta_{inf})
(1+\delta_{vr}+\delta_{vac})
+ \frac{\sigma_{hard}}{\sigma_0}.
\label{eq1}\end{equation}
Here $\delta_{vr}$ and $\delta_{vac}$ come from so-called virtual
corrections: the vertex function and vacuum polarization by lepton
pairs and
hadrons, resp. $\delta_{inf}$ appears after cancellation of infrared
divergences. The exponential results from the summation of soft photon
contributions
over all orders of perturbation theory. 
$\sigma_{hard}$ is the contribution of hard photon radiation.
In the 
general case it has the form of a three-dimensional integral over the
photon phase space. 
In Ref.\ \cite{Aku} the numerical
analysis was done for fixed target experiments. In this report
we provide numerical results also for deep inelastic scattering at HERA.
Moreover, we discuss the case of photoproduction within the 
semi-analytical approach. 

Another possible approach for the calculation of RC in semi-inclusive and
exclusive processes is based on using a Monte-Carlo generator for
inclusive DIS.
In this report we discuss how the RC can be calculated using
information obtained during event generation. Also we show that, under
certain assumptions, both methods give the same results. This fact is
illustrated for the case of measurements at HERMES.

It should be noted that all numerical results are shown for the electron
method of the reconstruction, when the kinematical variables are
calculated using the measurement of the final lepton. There are other
reconstruction methods (for example, the double angle method using for the
case of diffractive vector meson production), which are, in the most of
the phase space, less sensitive to these radiative corrections. 

\section{The code DIFFRAD}\label{diffrad}

The FORTRAN code DIFFRAD was developed to calculate radiative
corrections for diffractive vector meson electroproduction and is based
on the semi-analytical approach. This code has two versions which are
referred to as IDIFFRAD.F and MDIFFRAD.F. The first version allows to
calculate the RC to the cross section ($ d^4\sigma / dxdQ^2dtd\phi_h$
and $ d^3\sigma / dxdQ^2dt$) of diffractive vector meson
electroproduction using usual (non Monte-Carlo) methods of numerical
integration.  The second one exploits the same formulae
but using Monte Carlo methods for integration. It allows to integrate
not only over the photon kinematic variables, but also over those of the
final lepton.  Integration over $Q^2$, $W^2$ and $t$ in arbitrary bins
can be performed as well as over the full kinematically allowed phase
space of the final lepton. Therefore MDIFFRAD.F has more possibilities
in comparison to IDIFFRAD.F, but it requires more CPU time.

The code requires input for the model of the virtual photoproduction
cross sections $\sigma_{L,T}$ and allows to apply 
a cut on inelasticity.
As input for $\sigma_{L,T}$ we use the model presented originally in
Ref.\ \cite{Ryskin} and refined in Ref.\ \cite{DIPSI} ({DIPSI}). 
The implementation of other models is possible and
straightforward.  A cut on the inelasticity $v$ ($v=p_h^2-M^2$, where
$p_h$ is the total momentum of unobserved particles and $M$ is the
proton mass) is often suitable to reduce the RC.  The value of the cut
is related to the experimental resolution on $v$ and depends on details
of the experimental procedure of non-exclusive background subtraction. A
detailed
discussion of this point can be found in Ref.\ \cite{Aku}. Both versions
of DIFFRAD allow one to calculate RC with and without this cut. Note
that two additional options also exist for the calculation of the
inelasticity distribution with and without RC and for the fast
evaluation of an approximate calculation of $\sigma_{hard}$.

The FORTRAN code { DIFFRAD} can be received upon request via email from 
{\tt aku@hep.by}

\section{Electroproduction}\label{electrop}

\begin{figure}[h]
\unitlength 1mm
\begin{center}
\begin{picture}(180,120)
\put(25,122){\makebox(0,0){\small  $\eta$}}
\put(88,122){\makebox(0,0){\small  $\eta$}}
\put(25,66){\makebox(0,0){\small  $\eta$}}
\put(88,66){\makebox(0,0){\small  $\eta$}}
\put(56,115){\makebox(0,0){\scriptsize  $W^2=1000$GeV$^2$}}
\put(120,115){\makebox(0,0){\scriptsize  $W^2=4000$GeV$^2$}}
\put(56,59){\makebox(0,0){\scriptsize  $W^2=8000$GeV$^2$}}
\put(120,59){\makebox(0,0){\scriptsize  $W^2=12000$GeV$^2$}}
\put(130,70){\makebox(0,0){\scriptsize  $Q^2$,GeV$^2$}}
\put(65,70){\makebox(0,0){\scriptsize  $Q^2$,GeV$^2$}}
\put(130,15){\makebox(0,0){\scriptsize  $Q^2$,GeV$^2$}}
\put(65,15){\makebox(0,0){\scriptsize  $Q^2$,GeV$^2$}}
\put(5,0){
\epsfxsize=14cm
\epsfysize=14cm
\epsfbox{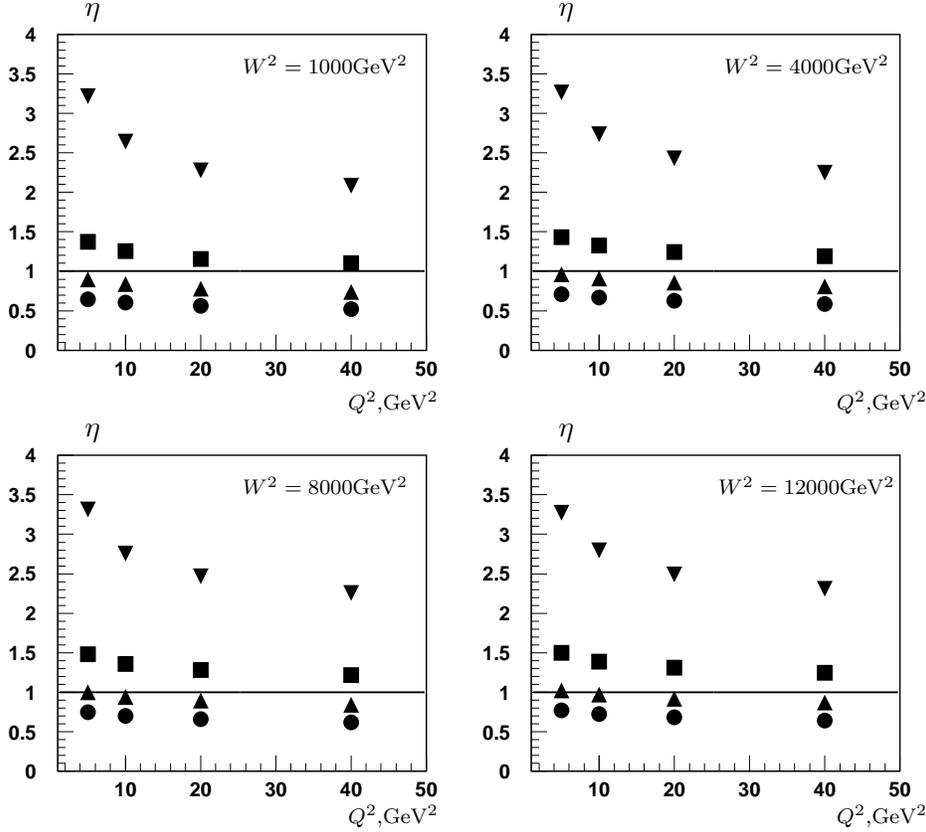}
}
\end{picture}
\caption{\label{h1wqqq}
RC factor for $\rho(770)$ electroproduction at HERA; $\sqrt{s}$=300
GeV.  
Symbols from top to bottom correspond to
$t=-0.7$, $-0.5$, $-0.3$, $-0.1$ GeV$^2$}
\end{center}
\end{figure}

\begin{figure}[t]\centering
\unitlength 1mm
\parbox{.4\textwidth}{\centering
\begin{picture}(80,80)
\put(65,5){\makebox(0,0){\small  $-t$,GeV$^2$}}
\put(55,40){\makebox(0,0){\bf observed}}
\put(45,25){\makebox(0,0){\bf Born}}
\put(15,72){\makebox(0,0){\small  $\sigma, mb$}}
\put(-5,0){
\epsfxsize=8cm
\epsfysize=8cm
\epsfbox{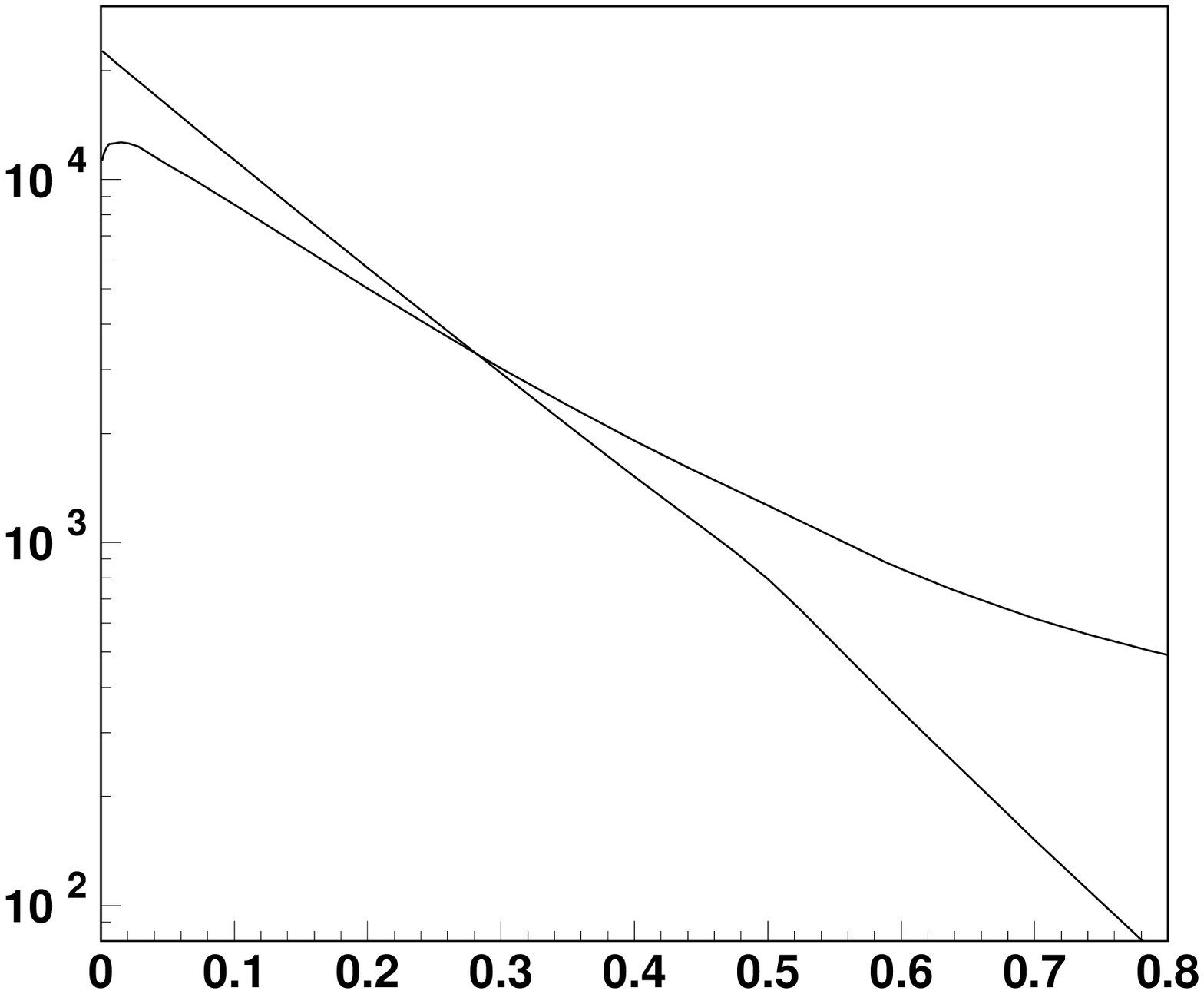}
}
\end{picture}
\caption{\label{tdep} Observable and Born cross sections
 for exclusive $\rho(770)$ electroproduction as a function of $t$;
$\sqrt{s}$=300
 GeV; $W$=70 GeV; 
$Q^2$=4
 GeV$^2$. 
} 
}
\hfill
\parbox{.4\textwidth}{\centering
\begin{picture}(80,80)
\put(60,5){\makebox(0,0){\small  $-t$,GeV$^2$}}
\put(10,72){\makebox(0,0){\small  $b$,GeV$^{-2}$}}
\put(-8,0){
\epsfxsize=8cm
\epsfysize=8cm
\epsfbox{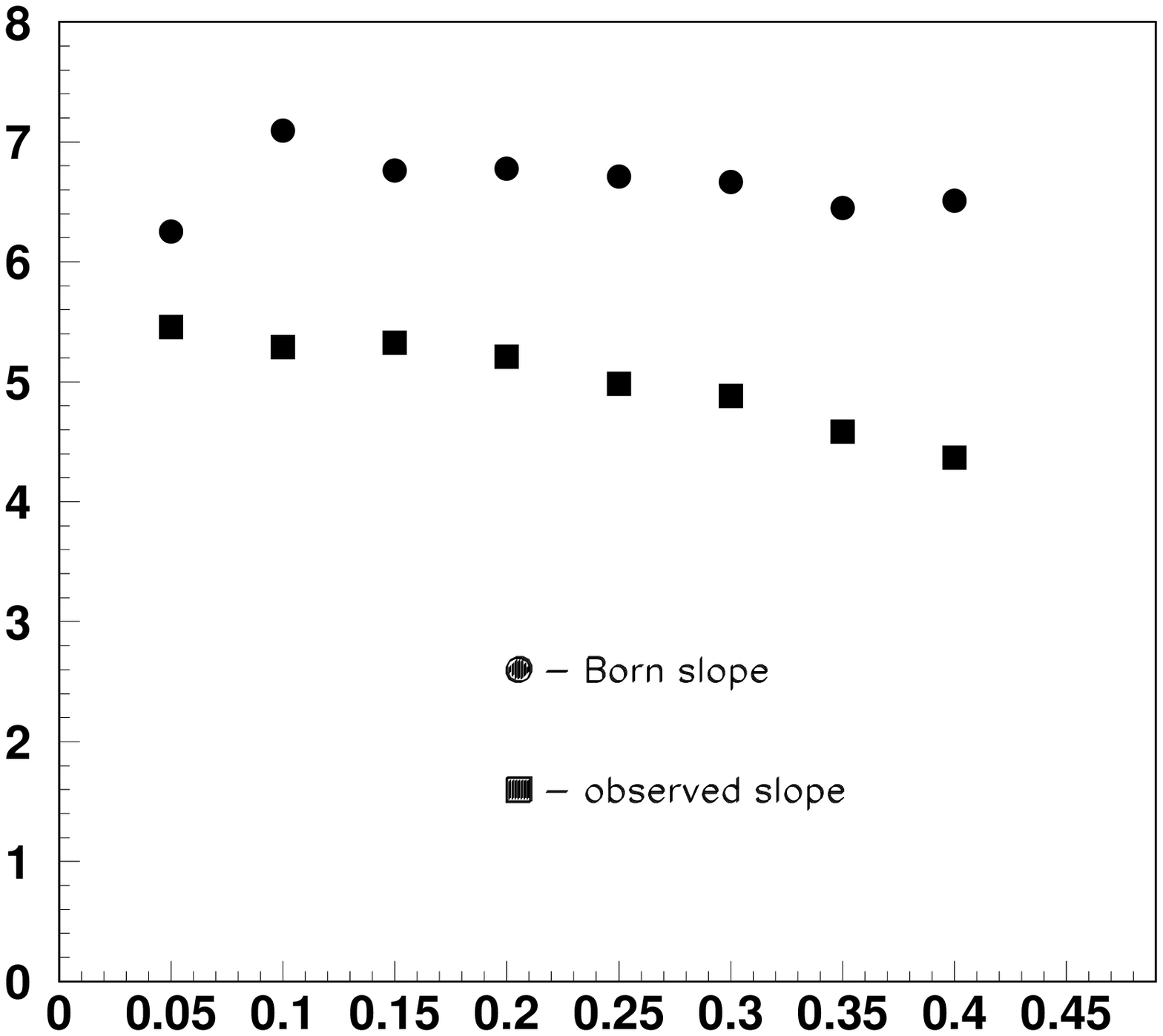}
}
\end{picture}
\caption{\label{slope} Slope parameter for $\rho(770)$ electroproduction; 
$\sqrt{s}$=300 GeV; $W$=70 GeV; $Q^2$=4
 GeV$^2$. 
}}
\end{figure}

Results obtained with the help of DIFFRAD for the three-fold
differential cross section are presented in Fig.\ \ref{h1wqqq}
($\sqrt{s}$=300~GeV).
Different plots in this figure show that there is no strong
$W$-dependence of the RC factor $\eta$. This is due to the fact that
$\sigma_{L}$ and $\sigma_{T}$ are almost flat in the considered
kinematical region. Moreover there is no large explicit dependence on
$W$ in the RC factor itself. The quantities $\delta_{vr}$ and
$\delta_{vac}$ in Eq.\ (\ref{eq1}) are practically independent of $W$
and only a small dependence comes from $\delta_{inf}$ and
$\sigma_{hard}$.  The $Q^2$-dependence shown in this figure is typical
for the case when the inelasticity cut is not applied. If this cut is
used then the rise of $\eta$ when $Q^2$ goes down would be
suppressed. From this figure one can also see that the $t$-dependence is
rather important. Figures \ref{tdep} and \ref{slope} illustrate this
last property. $\eta$ crosses unity for $-t \sim $ 0.25--0.3
GeV$^2$ and rises with increasing $|t|$. The large positive correction
in this case is a result of the large phase space for photon radiation.
The cut applied to the inelasticity (or $E-p_z$) can again reduce the
value of the RC factor for large $|t|$. As a consequence of the
$t$-dependence of $\eta$, the observed slope parameter also receives
large RC.  This is illustrated in Fig.\ \ref{slope}.  The Born cross
section has a steeper slope with respect to $|t|$ than the observed
cross section. The radiative correction to the slope parameter is
negative and $ \sim 10\%$.

\begin{figure}[t]\centering
\unitlength 1mm
\parbox{.4\textwidth}{\centering
\begin{picture}(80,80)
\put(10,72){\makebox(0,0){$\eta$}}
\put(55,62){\makebox(0,0){\small  $-t$=0.3GeV$^2$}}
\put(58,40){\makebox(0,0){\small  $-t$=0.2GeV$^2$}}
\put(55,18){\makebox(0,0){\small  $-t$=0.1GeV$^2$}}
\put(65,5){\makebox(0,0){\small  $\phi_h$}}
\put(-5,0){
\epsfxsize=8cm
\epsfysize=8cm
\epsfbox{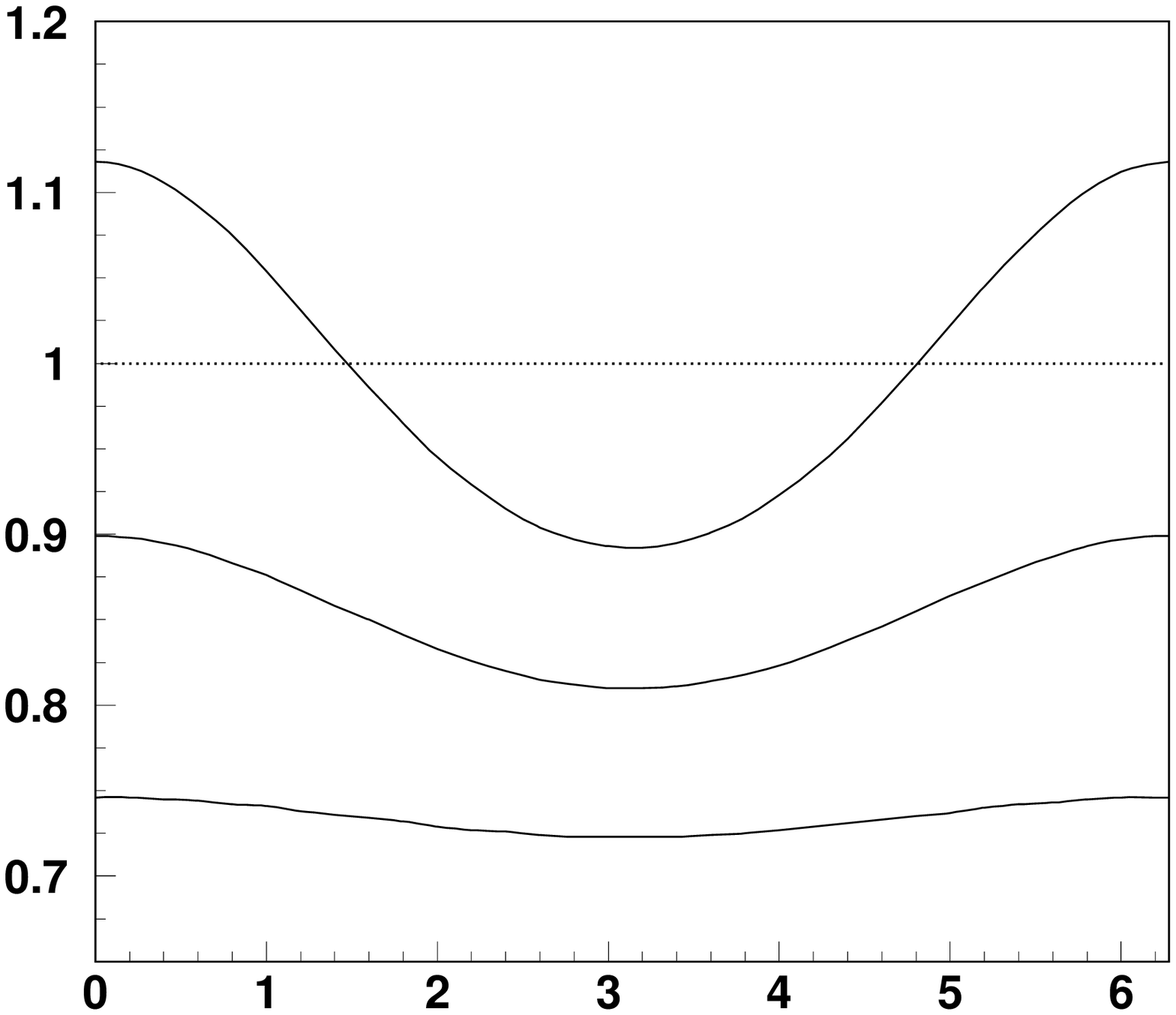}
}
\end{picture}
\caption{\label{phidep} RC factor for exclusive $\rho(770)$
electroproduction as a 
  function of 
$\phi_h$; $\sqrt{s}$=300 GeV; $W$=70 GeV. 
} 
}
\hfill
\parbox{.4\textwidth}{\centering
\begin{picture}(80,80)
\put(55,5){\makebox(0,0){$-t$,GeV$^2$}}
\put(10,72){\makebox(0,0){$\eta$}}
\put(18,65){\makebox(0,0){\small $W$=30 GeV}}
\put(18,60.4){\makebox(0,0){\small $W$=50 GeV}}
\put(18,55.8){\makebox(0,0){\small $W$=70 GeV}}
\put(18,51.2){\makebox(0,0){\small $W$=90 GeV}}
%\put(58,55){\makebox(0,0){\small   \bf No cut on $W^2$ }}
%\put(50,30){\makebox(0,0){ \small   \bf 60$<W<80$ GeV }}
\put(-8,0){
\epsfxsize=8cm
\epsfysize=8cm
\epsfbox{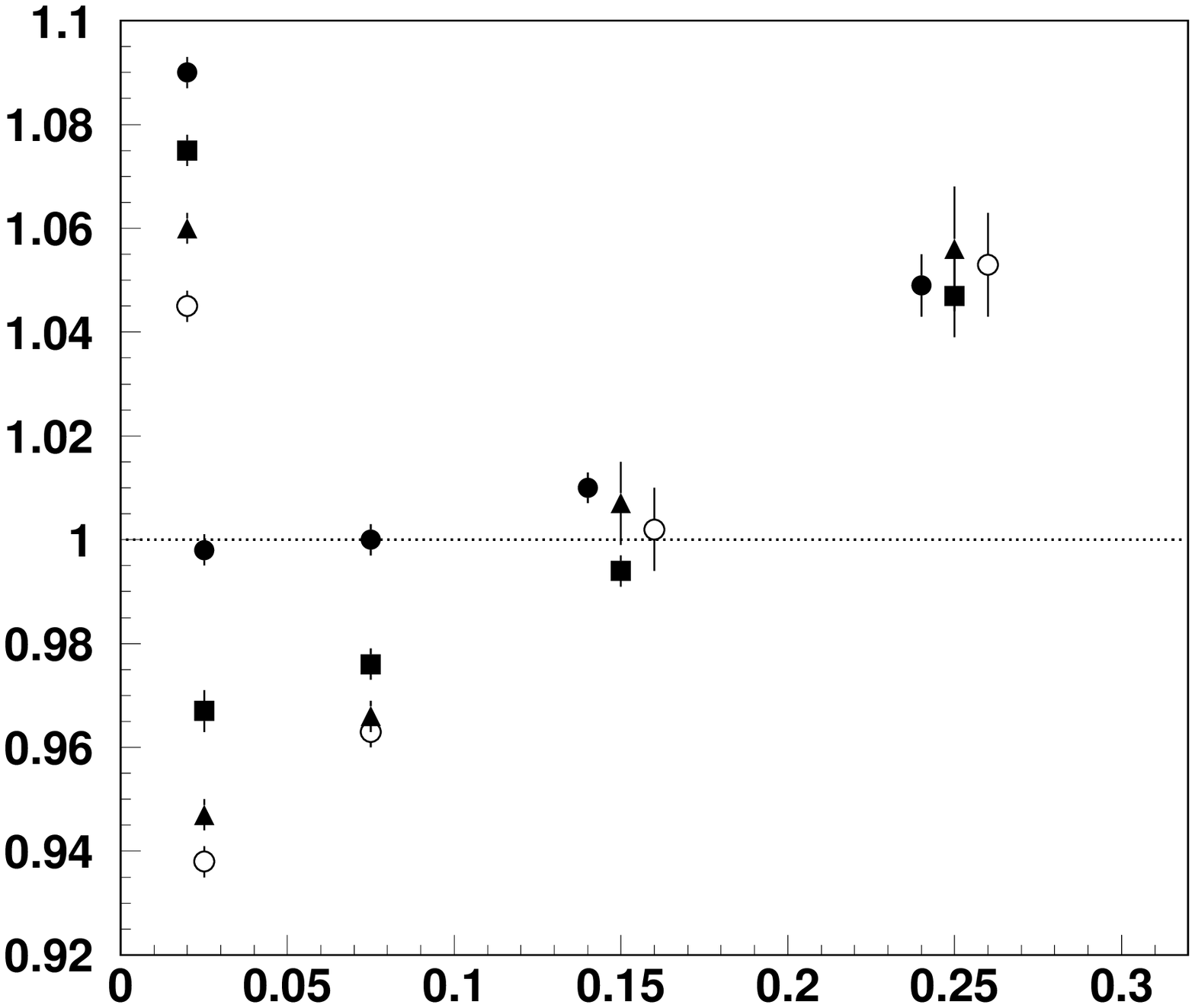}
}
\end{picture}
\caption{\label{photo} RC factor for exclusive $\rho(770)$ photoproduction
as a
  function of t;
$\sqrt{s}$=300 GeV.
}}
\end{figure}

The model used for exclusive virtual photoproduction of vector mesons is
based on
the hypothesis of s-channel helicity conservation (SCHC)\footnote{We
  should note that now there are experimental indications for the
  violation of this hypothesis \cite{ZEUS,H1} as well as theoretical
  models beyond SCHC \cite{IK,KNZ}.  Unfortunately the development of
  DIFFRAD for using such models is not straightforward, because the
  calculation of additional terms is required.  The calculation for the
  general case of scalar meson production is considered in \cite{ASSh}.}.
In the case of SCHC the Born cross section is independent of the
azimuthal angle $\phi_h$ between the scattering and the production
plane. However, the observed cross section does depend on $\phi_h$,
because the formulae for RC includes this angle. Figure \ref{phidep}
shows an example.  Since only $\sigma_{hard}$ has an essential
dependence on the azimuth, $\eta$ has a visible $\phi_h$ dependence at
large $|t|$, where $\sigma_{hard}$ gives a relatively large contribution
to the total RC.

We note finally that the RC to the cross section of diffractive vector
meson electroproduction is very sensitive to the inelasticity cut.  This
cut suppresses the contribution of hard photon radiation, which is
always positive.  Thus the use of harder cuts leads to smaller values of
the radiative correction factor. However, in the region of small $|t|$
where radiative corrections are negative ($\eta < 1$) the influence of
the cut is not essential or leads to larger absolute values of RC
($\eta$-1). We do not apply the cut for this numerical analysis.

All numerical results shown here were obtained for the case of
exclusive $\rho(770)$
production. However there
is no 
essential dependence on the type of the observed vector meson, and all
discussed
features are very similar for the production of heavier vector
mesons.

\section{Photoproduction}\label{photop}

In photoproduction interactions, i.e.\ very small $Q^2$ (quasi-real
photon exchanged), the final positron is scattered at very small angle and
escapes detection in the main detector through the beam pipe.
In this case QED
radiative corrections can be calculated by integration of the analytical
formulae over the phase space of the
final positron. A numerical analysis shows
that radiative corrections in this case are smaller than in the case of
electroproduction. The results are shown in Fig.\ \ref{photo}. 
%Here the
%integration over the full kinematically allowed region of $Q^2$ was
%performed for several different values of $W^2$.  
Note that small values
of RC in this case reflect the statement of the Kinoshita-Lee-Nauenberg
theorem \cite{Kino,LeeN}. For our case it says that if we integrate
over the phase space of a particle which radiates, then all leading
logarithms are mutually cancelled (see Ref.\ \cite{KU}, for example).

\section{Monte Carlo approach}\label{MCapproach}

\begin{figure}[t]\centering
\unitlength 1mm
\parbox{.4\textwidth}{\centering
\begin{picture}(80,80)
\put(5,0){
\epsfxsize=6cm
\epsfysize=8cm
\epsfbox{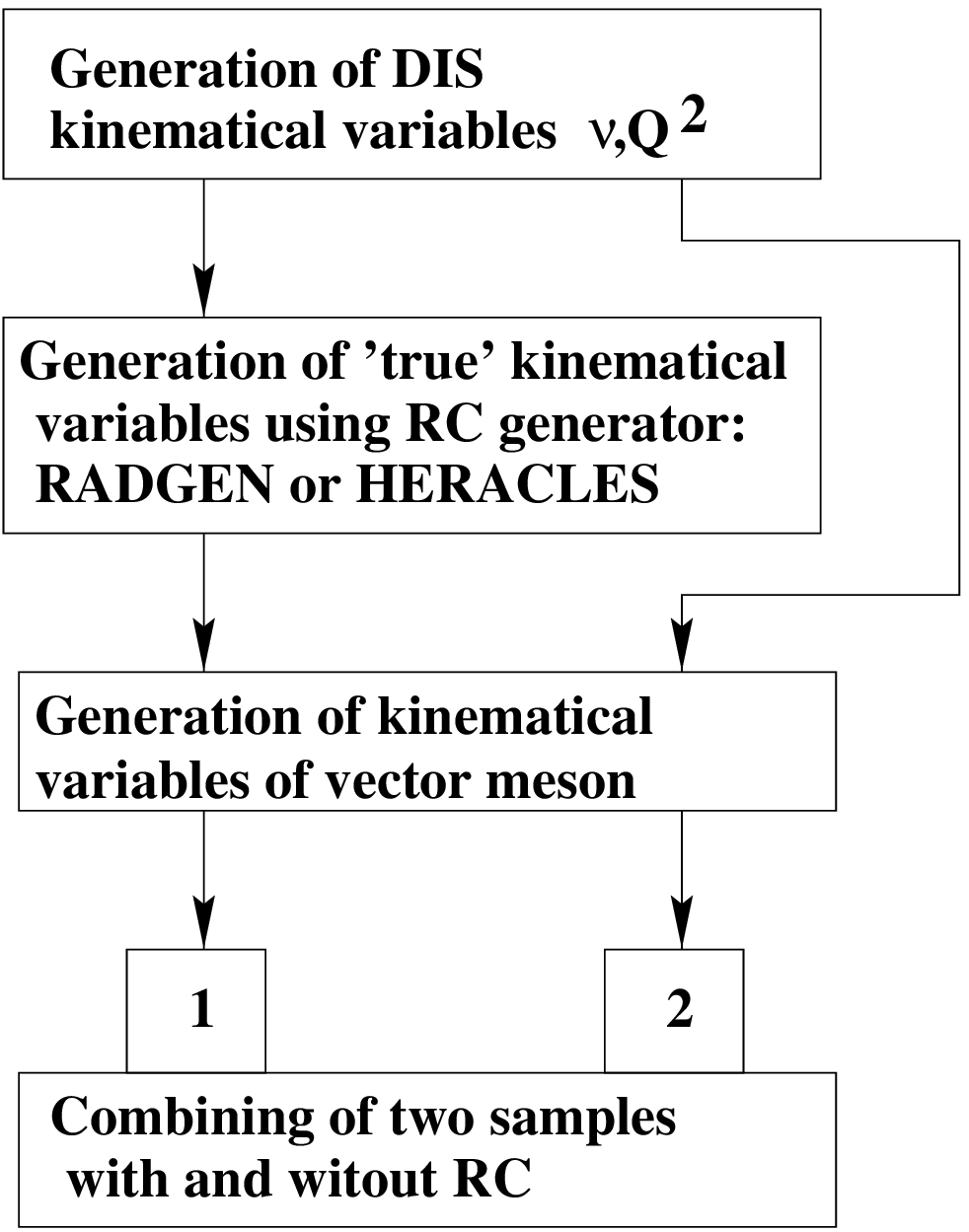}
}
\end{picture}
\caption{\label{scheme}Possible scheme of Monte Carlo calculation of the 
  RC factor.} 
}
\hfill
\parbox{.4\textwidth}{\centering
\begin{picture}(80,80)
\put(55,5){\makebox(0,0){$-t$,GeV$^2$}}
\put(10,72){\makebox(0,0){$\eta$}}
\put(-8,0){
\epsfxsize=8cm
\epsfysize=8cm
\epsfbox{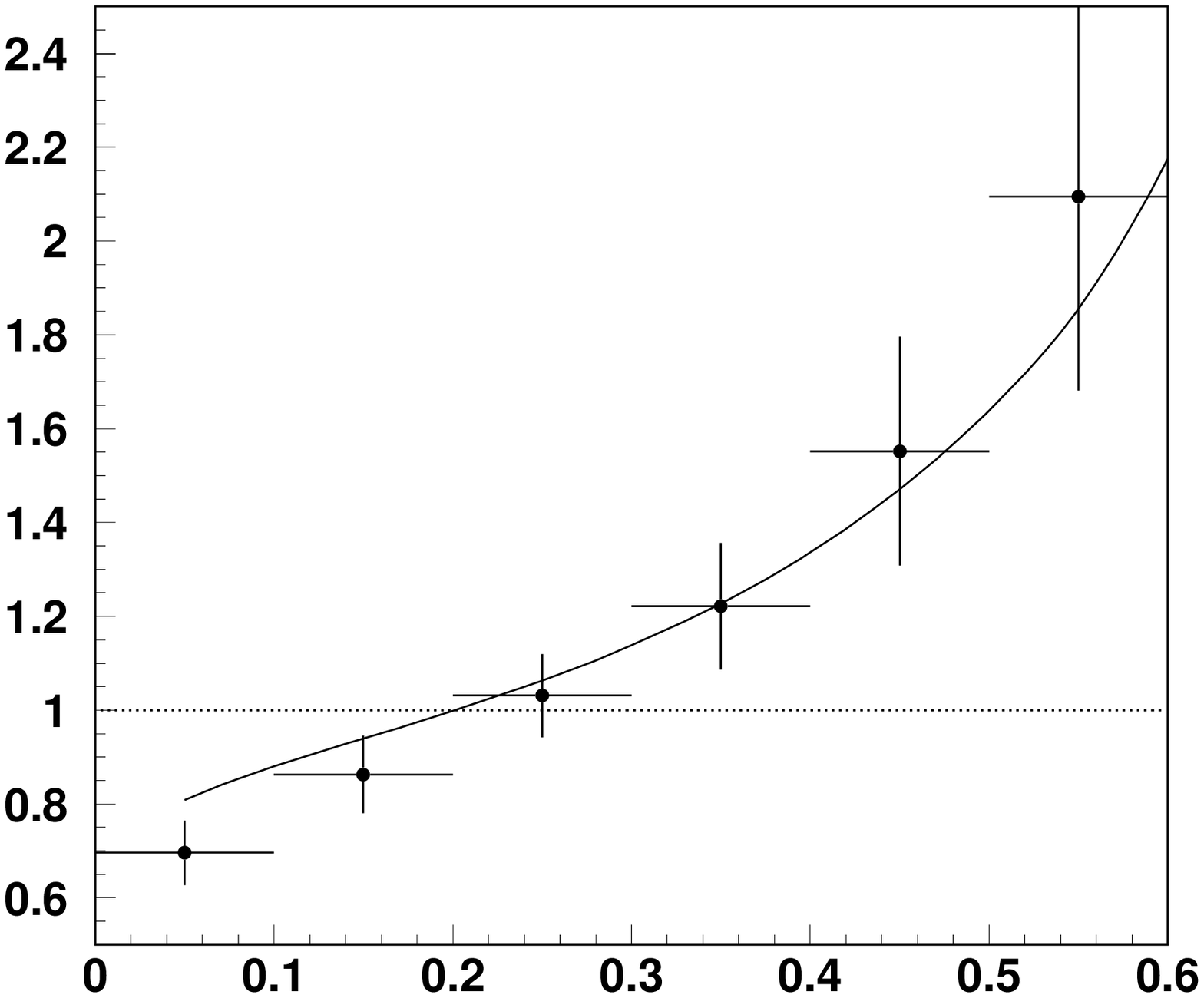}
}
\end{picture}
\caption{\label{mcappr} RC factor for HERMES from DIFFRAD (semi-analytic 
  approach, full line) and from RADGEN (Monte Carlo approach, points
  with error bars); $\sqrt{s}$=7.18 GeV; $Q^2$=1.5 GeV$^2$; $W^2$=20
  GeV$^2$.}} 
\end{figure}

The Monte Carlo approach considered here  is based on a 
generator for inclusive processes with radiative effects included.
Here we focus on the case of HERMES and use RADGEN
\cite{RADGEN} 
as such a generator. This generator was developed for the
fixed target experiments while for HERA experiments 
 HERACLES \cite{HERACLES} which includes electroweak effects should be
used.

Two Monte Carlo samples with and without radiative effects have to be
prepared.  The procedure is sketched in Fig.\ \ref{scheme}. Firstly, the
kinematical variables $\nu$ and $Q^2$ defining the momentum of the final
lepton are generated in accordance with the Born cross section.  In the
case of generation of the sample without RC these variables are used as
an input to generate the momentum of the vector meson.  For the
generation of the sample with RC, a Monte Carlo generator including real
photon emission has to be used. Three main tasks are performed by this
generator:
\begin{enumerate}
\item simulation of the appropriate scattering channel (non-radiative;
  elastic, quasielastic or inelastic radiative tails) in accordance with
  their contribution to the total observed cross section;
\item calculation of the inclusive RC factor $\eta^{inc}$ for the given
  values of $\nu$ and $Q^2$. Since for both samples we generate the same
  number of events this factor has to be used as a weight to
  obtain the observed cross section as a weighted sum of
  events\footnote{Another algorithm without calculation of weights is
    possible. In this case both samples with and without RC are
    generated in accordance with the observed and the Born cross
    section, resp.  Both approaches are equivalent, and everything
    discussed below is valid for this case as well.};
\item generation of the kinematical variables $\nu_{true}$ and
  $Q^2_{true}$ which are then used as input for generating the momentum
  of the vector meson. Note that $\nu_{true}$ and $Q^2_{true}$ differ
  from $\nu$, $Q^2$ for events which contain a radiated photon.
\end{enumerate}
The events are collected in predefined bins.  If the numbers of events
in a given bin for the samples with and without radiative corrections
are $n_1$ and $n_2$, resp., the RC factor for vector meson
electroproduction $\eta_{\rho}$ can be estimated as
\begin{equation}
\eta_{\rho}=\frac{n_1<\;\eta^{inc}\;>}{n_2}
\end{equation}
where $\eta^{inc}$ is the RC factor for the inclusive process.  We note
that due to photon radiation $\nu_{true} \leq \nu$. Thus the
exclusive production
of a vector meson with a given value for its energy (which is related to
$|t|$) is less probable for a radiative event than for a non-radiative
event with the same meson energy. As a result $n_1<n_2$ and the typical
value of the RC factor for vector meson production (for semi-inclusive
processes in general) is smaller than for the inclusive case.

Figure \ref{mcappr} presents results for the HERMES experiment obtained
within the two approaches. From this picture we can conclude that the
Monte Carlo approach correctly reproduces the RC factor for vector meson
electroproduction. The inelasticity cut was not applied for these
calculations.
%It is easy to obtain  it within the semi-analytical approach,
%however
%further development is required for considered Monte Carlo scheme. 
We note that the parameter $\Delta$ which separates soft from hard
photon radiation and which is always necessary in the Monte Carlo
approach (see discussion in Ref.\ \cite{RADGEN}) should be chosen
smaller than its default value $\Delta=100$MeV. This value is close to
the threshold energy for observing photons in the HERMES detector.  
In
the case of vector meson electroproduction 
the maximal photon energy is smaller (especially for small $|t|$) in
comparison with the inclusive
case. So the parameter $\Delta$ has to be reduced also.
For the
present calculation $\Delta=5$MeV was chosen. For smaller values of
$\Delta$ we do not see an essential dependence on $\Delta$.

\section*{Acknowledgments}
I am grateful to A.Brull and N.Shumeiko for help and support. Also
I would like to thank N.Akopov, A.Borissov, S.Chekanov,
L.Favart, E.Kuraev, P.Kuzhir,
A.Nagaitsev, N.Nikolaev, A.Prosku\-ry\-kov, M.Ryskin,
and
H.Spiesberger for
fruitful discussions and comments.

\end{document}